\documentclass[twocolumn, twocolappendix]{aastex631}

\usepackage{newtxmath}
\usepackage{amsmath, graphics, bm, graphicx}

\allowdisplaybreaks

\newcommand{\hu}{{\bm {\hat u}}}
\newcommand{\hv}{{\bm {\hat v}}}
\newcommand{\hl}{{\bm {\hat l}}}
\newcommand{\hx}{{\bm {\hat x}}}
\newcommand{\hy}{{\bm {\hat y}}}
\newcommand{\hz}{{\bm {\hat z}}}
\newcommand{\hn}{{\bm {\hat n}}}
\newcommand{\btimes}{{\bm \times}}
\newcommand{\bcdot}{{\bm \cdot}}

\newcommand{\vn}{{\bm n}}
\newcommand{\der}{{\rm d}}
\newcommand{\appropto}{\mathrel{\vcenter{
  \offinterlineskip\halign{\hfil$##$\cr
    \propto\cr\noalign{\kern2pt}\sim\cr\noalign{\kern-2pt}}}}}

\begin{document}

\title{Concealing Circumbinary Planets with Tidal Shrinkage}

\author[0009-0003-9522-7004]{Saahit Mogan}\thanks{smogan@berkeley.edu}
\affiliation{Department of Astronomy, 
University of California, Berkeley, 
Berkeley, CA 94720, USA}

\author[0000-0002-9849-5886]{J.J. Zanazzi}\thanks{51 Pegasi b fellow}
\affiliation{Department of Astronomy, 
University of California, Berkeley, 
Berkeley, CA 94720, USA}

\begin{abstract}

Of the 14 transiting planets that have been detected orbiting eclipsing binaries (`circumbinary planets'), none have been detected with stellar binary orbital periods shorter than 7 days, despite such binaries existing in abundance.  The eccentricity-period data for stellar binaries indicates that short-period ($< 7$ day) binaries have had their orbits tidally circularized. 
We examine here to what extent tidal circularization and shrinkage can conceal circumbinary planets, i.e.~whether planets actually exist around short-period binaries, but are not detected because their transit probabilities drop as tides shrink the binary away from the planet. We carry out a population synthesis by initializing a population of eccentric stellar binaries hosting circumbinary planets, and then circularizing and tightening the host orbits using stellar tides. 
To match the circumbinary transit statistics, 
stellar binaries must form with eccentricities $\gtrsim$ 0.2 and periods $\gtrsim$ 6 days, with circumbinary planets emplaced on exterior stable orbits before tidal circularization; moreover, tidal dissipation must be efficient enough to circularize and shrink binaries out to $\sim$6--8 days. The resultant binaries that shrink to sub-7-day periods no longer host transiting planets.
However, this scenario cannot explain the formation of nearly circular, tight binaries, brought to their present sub-seven-day orbits from other processes like disk migration.
Still, tidal shrinkage can introduce a bias against finding transiting circumbinary planets, and predicts a population of KIC 3853259 (AB)b analogs consisting of wide-separation, non-transiting planets orbiting tight binaries.
\end{abstract}

\keywords{}

\section{Introduction} \label{sec:intro}

At least 32 sub-stellar companions orbiting stellar or brown-dwarf binaries have been detected to date.  Nine companions were directly imaged \citep[e.g.][]{Burgasser+(2010), Kuzuhara+(2011), Delorme+(2013), Bryan+(2016b), Bonavita+(2017), Dupuy+(2018), Dupuy+(2023), Janson+(2019), Janson+(2021)}, five were discovered by microlensing \citep[e.g.][]{Bennett+(2016), Han+(2017), Han+(2020), Han+(2024), Kuang+(2022)}, two by radial velocity variations (e.g. \citealt{Correia+(2005), Standing+(2023)}, see also \citealt{Goldberg+(2023)}), and two from timing variations in eclipsing binaries \citep[e.g.][]{Borkovits+(2016), Getley+(2017)}.  Fourteen were discovered when the planet transited an eclipsing binary host, using \textit{Kepler} \citep{Doyle+(2011), Welsh+(2012), Welsh+(2015), Orosz+(2012a), Orosz+(2012b), Orosz+(2019), Schwamb+(2013), Kostov+(2014), Kostov+(2016), Socia+(2020)} and TESS \citep{Kostov+(2020), Kostov+(2021)} mission photometry.  The detection of transiting circumbinary planets favors planets that are nearly co-planar to their binaries; every transiting circumbinary planet is inclined by only a few degrees to the binary's orbital plane. 

Low-inclination circumbinary planets form from  low-inclination protoplanetary disks. For binaries with periods $P_b \lesssim 100$ days, circumbinary disks are observed to be nearly co-planar \citep{Czekala+(2019)}. Viscous dissipation usually works to damp a circumbinary disk's inclination \cite[e.g.][]{FoucartLai(2013), FoucartLai(2014), Moody+(2019), Smallwood+(2019)}. Larger binary eccentricities allow for more misaligned disks and by extension more misaligned planets (e.g.~\citealt{Aly+(2015), MartinLubow(2017), ZanazziLai(2018), LubowMartin(2018), CuelloGiuppone(2019), Czekala+(2019)}).  KH-15D \citep[e.g.][]{ChiangMurrayClay(2004), Winn+(2004), Winn+(2006), Poon+(2021)} and Bernhard-2 \citep{Zhu+(2022), Hu+(2024)} are precessing, nearly-aligned disks which occult their host binaries, which may be progenitors of transiting circumbinary planet systems.

\begin{figure}
    \includegraphics[width=\linewidth]{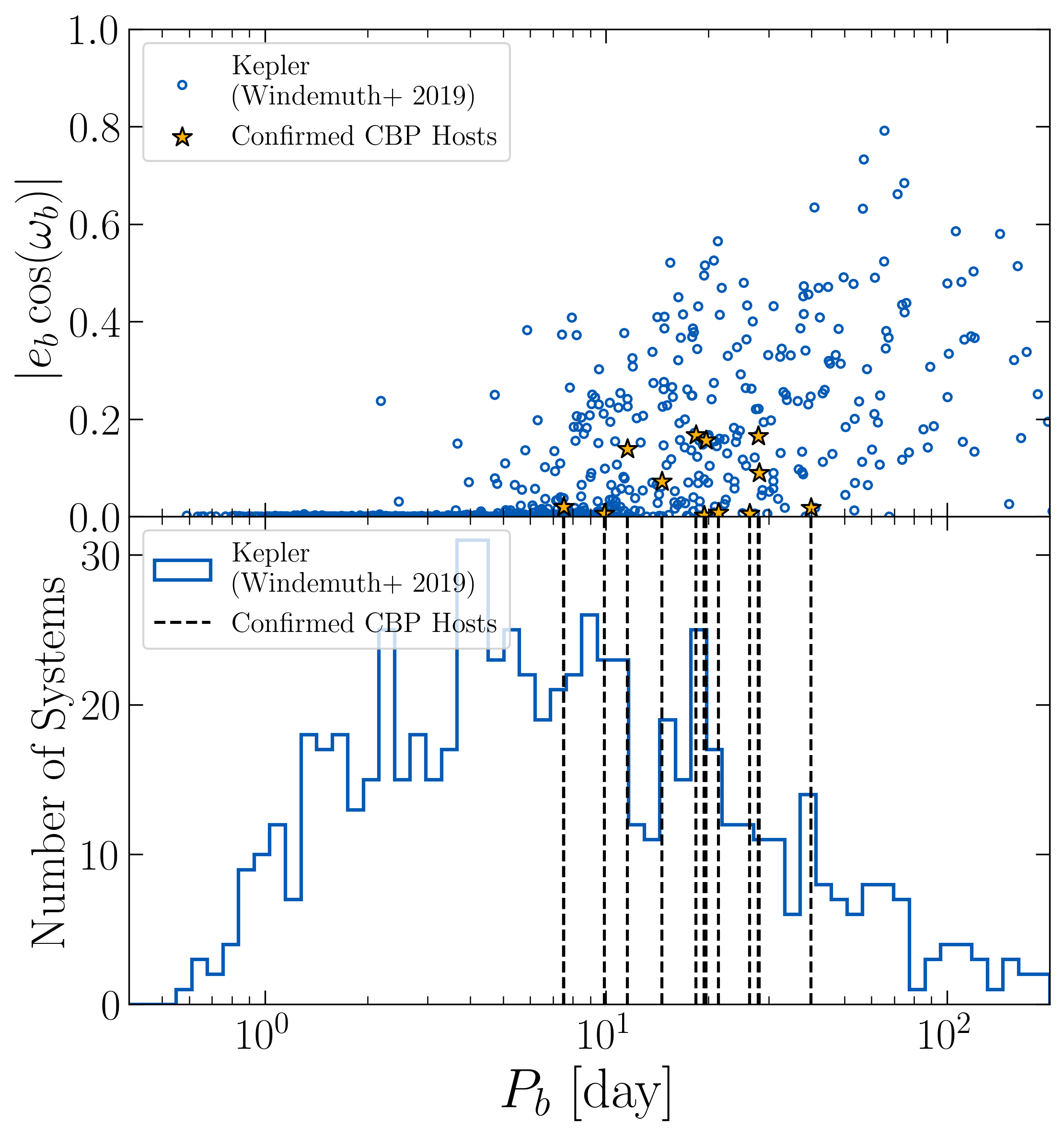}
    \caption{
    Comparing the \textit{Kepler} eclipsing binaries \citep{Windemuth+(2019)} to binaries hosting transiting circumbinary planets (see Table~\ref{tbl:obs}), varying the binary period $P_b$.  \textit{Top panel}: Component of the binary eccentricity projected onto the observer's sky plane. Measurement errors are typically $<1\%$. \textit{Bottom panel}: Histogram of $P_b$ values for \textit{Kepler} eclipsing binaries, with vertical lines marking binaries which host planets. Planets are not detected when $P_b < 7 \ {\rm days}$, where most binaries have nearly-circular orbits.
    \label{fig:motiv}
    }
\end{figure}

\begin{table*}
    \begin{center}
    \begin{tabular}{|c|c|c|c|c|c|c|c|}
        \hline
        Planet Name & $M_A\ [M_\odot]$ & $M_B\ [M_\odot]$ 
        & $P_b\ [\rm day]$ & $P_p\ [day]$ & $e_b$ & $|e_b\cos \omega_b|$ & Reference \\ 
        \hline
        Kepler 16b        & $0.6897$ & $0.2026$ & $39.95$ & $226.75$  & $0.1594$ & $0.0181$ & \cite{Doyle+(2011)} \\
        Kepler 34b        & $1.0479$ & $1.0208$ & $28.00$ & $288.9$   & $0.5209$ & $0.1658$ & \cite{Welsh+(2012)} \\
        Kepler 35b        & $0.8877$ & $0.8094$ & $21.39$ & $130.20$  & $0.1421$ & $0.0086$ & \cite{Welsh+(2012)} \\
        Kepler 38b        & $0.949$  & $0.249$  & $19.37$ & $104.03$  & $0.1032$ & $0.0024$ & \cite{Orosz+(2012b)} \\
        Kepler 47b        & $1.043$  & $0.362$  & $7.52$  & $50.72$   & $0.0234$ & $0.0199$ & \cite{Orosz+(2012a)} \\
        Kepler 47c$^*$    & "        & "        & "       & $304.08$  & "        & "        & \cite{Orosz+(2012a)} \\
        Kepler 47d$^*$    & "        & "        & "       & $180.41$  & "        & "        & \cite{Orosz+(2019)} \\
        Kepler 64b (PH1b) & $1.384$  & $0.386$  & $18.38$ & $131.13$  & $0.2117$ & $0.1677$ & \cite{Schwamb+(2013)} \\
        Kepler 413b       & $0.820$  & $0.5423$ & $9.90$  & $67.65$   & $0.0365$ & $0.0062$ & \cite{Kostov+(2014)} \\
        Kepler 453b       & $0.944$  & $0.1951$ & $26.36$ & $242.34$  & $0.0524$ & $0.0063$ & \cite{Welsh+(2015)} \\
        Kepler 1647b$^*$  & $1.2207$ & $0.9678$ & $11.57$ & $1107.21$ & $0.1602$ & $0.1386$ & \cite{Kostov+(2016)} \\
        Kepler 1661b      & $0.841$  & $0.262$  & $28.16$ & $175.39$  & $0.112$  & $0.0901$ & \cite{Socia+(2020)} \\
        TOI 1338b         & $1.038$  & $0.3168$ & $14.62$ & $91.61$   & $0.1560$ & $0.0722$ & \cite{Kostov+(2020)} \\
        TIC 172900988b    & $1.238$  & $1.202$  & $19.66$ & $200.45$  & $0.4479$ & $0.1561$ & \cite{Kostov+(2021)} \\
        \hline
    \end{tabular}
    \end{center}
    \caption{Our sample of transiting circumbinary planets. \cite{MartinTriaud(2014), Li+(2016)} argued that systems marked with $*$ were statistical outliers.  For TIC 172900988b, we choose the Family 5 solution \citep{Kostov+(2021)}.
    \label{tbl:obs}}
\end{table*}

The farther removed a circumbinary planet is from its binary host, the more difficult it is to detect the planet from stellar transits, because the range of planet-binary inclinations which lead to transits is smaller.  If the formation statistics of planets is independent of the binary properties, observational selection effects favor the detection of planets around short-period binaries \citep[e.g.][]{Schneider(1994), MartinTriaud(2014), Li+(2016)}.  Surprisingly, no planets have been detected around sub-seven-day period binaries, despite most of the \textit{Kepler} eclipsing binaries having periods below seven days (Fig.~\ref{fig:motiv} bottom panel).  Several ideas have been put forth to explain why short-period binaries disfavor the formation of circumbinary planets. Binaries with $P_b < 10$ days might form through high-eccentricity migration, induced for example by the von Zipel-Lidov-Kozai (vZLK) effect from an inclined tertiary companion \citep[e.g.][]{MazehShaham(1979), FabryckyTremaine(2007), NaozFabrycky(2014), MoeKratter(2018)}.  Planets orbiting binaries undergoing vZLK cycles are engulfed, ejected, or placed on distant orbits misaligned to their host binaries, making them difficult to detect \citep[e.g.][]{MunozLai(2015), Martin+(2015), Hamers+(2016)}.
Another form of high-eccentricity migration involves a more nearly co-planar stellar triple, newly fragmented from a protostellar core/disk; the inner binary is kicked by the outer tertiary onto a high-eccentricity, low-periastron orbit, which subsequently circularizes and shrinks via 
energy dissipation from the disk \citep[e.g.][]{Bate+(2002), Bate(2009), Bate(2019)}. \cite{MoeKratter(2018)} argue that this channel (and not vZLK oscillations) dominates the formation of stellar binaries with $P_b < 10$ days, and again we would expect that planets in the vicinity of the inner binary as it circularizes would be consumed or expelled. Other dynamical instabilities could contribute to ridding short-period binaries of close-in planets as well, such as an evection resonance with an external perturber \citep{XuLai(2016)}, or the outward-then-inward migration of the binary caused by tidal synchronization followed by magnetic braking \citep{Fleming+(2018)}. 

\begin{figure}
    \includegraphics[width=\linewidth]{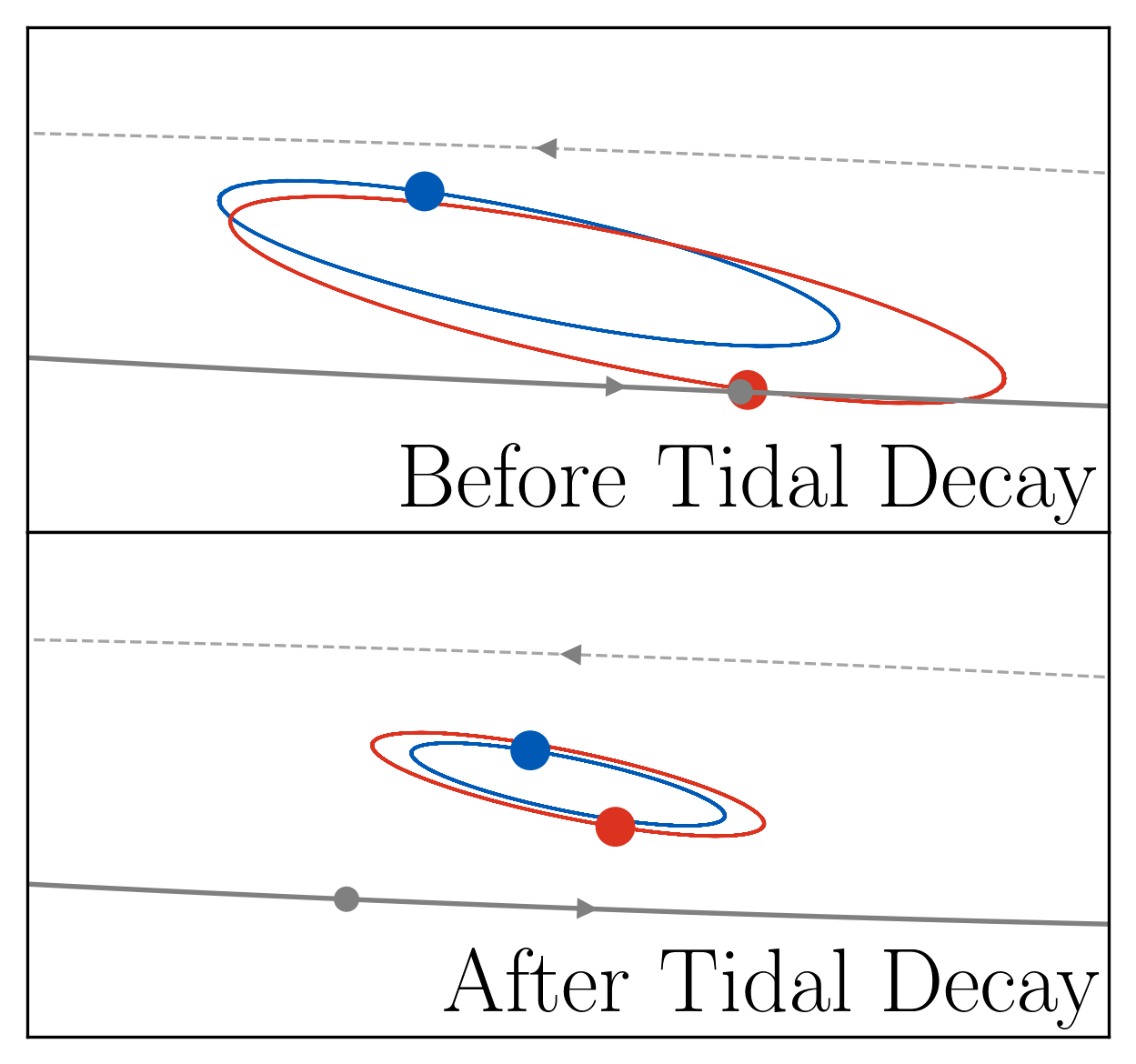}
    \caption{
    How binary circularization might worsen the chances of detecting transiting planets they host, as viewed by a distant observer.  For illustrative clarity, we display binaries which are not eclipsing. \textit{Top panel}: When the planet forms prior to orbital decay, it can cross the binary orbital plane.
    \textit{Bottom panel}: After orbital decay caused by tidal circularization, the planet no longer transits either component of the binary.
    \label{fig:bias}}
\end{figure}

Interestingly, the $P_b \approx 7$ day cutoff for binaries hosting transiting planets lies squarely within an empirical range of periods, quoted anywhere from $\sim$3 to $\sim$15 days, where tidal dissipation within stars circularizes binary orbits \citep[e.g.][]{MeibomMathieu(2005), Milliman+(2014), Triaud+(2017), JustesenAlbrecht(2021), Zanazzi(2022), Bashi+(2023), IJspeert+(2024)}.
Comparing the eccentricities of circumbinary planet hosts to those of the broader \textit{Kepler} eclipsing binary population, we see the shortest host period lies conspicuously close to the period that binaries transition from eccentric to circular (Fig.~\ref{fig:motiv} top panel).  As orbits circularize, their semi-major axes shorten. This leads to the central question of this paper: 
how much of the observed deficit of transiting planets around short-period binaries can be explained by orbital decay from tidal circularization? From Figure \ref{fig:bias}, we see that binary shrinkage can conceal planets by preventing them from transiting their binary hosts.

In this work, we calculate under what circumstances binary shrinkage from stellar eccentricity tides, by itself, can stop planets from transiting and thereby reproduce the circumbinary transit statistics in Fig.~\ref{fig:motiv}, obviating the need for dynamical instabilities to eject planets.  We acknowledge at the outset that tidal shrinkage alone cannot be the whole story, because there may be ways of forming planets around tight binaries where stellar tides play no role. For example, binaries may achieve nearly circular, $P_b < 10$ day periods from protostellar disk 
torques \citep[e.g.][]{GoldreichTremaine(1980), Bate+(2002), Bate(2009), Bate(2019), MoeKratter(2018), TokovininMoe(2020)}. The same disk that tightens and circularizes the binary could also, afterward, spawn a planet sufficiently close-in to transit.  Circularization cannot explain why when transiting planets are detected, their binary host tends to be more circular than their planet-less counterparts (compare eccentricities of stars to circles in Fig.~\ref{fig:motiv}).  Stellar eccentricity tides also have no apparent answer to the question of why most circumbinary planets have radii between $\sim$0.3--0.8 Jupiter radii \citep[see e.g. Table 1 of][]{Li+(2016)}.   Circumbinary planets appear to be mostly sub-Saturns, which around single stars are much rarer than sub-Neptunes. Recent work posits sub-Neptunes might be ejected by dynamical instabilities while planets migrate through the disk \citep{MartinFitzmaurice(2022), Fitzmaurice+(2022)}.

Although we cannot address all these mysteries, we suspect that tidal shrinkage does introduce a real bias against detecting transiting circumbinary planets, and we explore here how sensitive the transit statistics are to tidal dissipation parameters and initial conditions. It is also worth highlighting that the scenario of tidal shrinkage predicts that $P_b < 7$ day binaries are, in fact, orbited by planets, just not ones that transit---these may be detected by other means, e.g. timing variations of eclipsing binaries \citep{Borkovits+(2016)}. 
Section~\ref{sec:model} lays out our transiting circumbinary planet population synthesis model, including our model for orbital decay from tidal dissipation.  Section~\ref{sec:results} compares the results of our population synthesis to the observed circumbinary planets.  Section~\ref{sec:disc} summarizes and discusses.

\section{Transiting Circumbinary Planet Population Synthesis} \label{sec:model}

We construct a population synthesis model to test the effect tidal shrinkage has on the detectability of transiting circumbinary planets.  Our model consists of three parts: the orbital decay of stellar binaries, the distribution of circumbinary planets after formation, and an algorithm for determining which circumbinary planets transit.

\subsection{Binary Tidal Circularization} \label{subsec:binary}

\begin{figure*}
    \includegraphics[width=\linewidth]{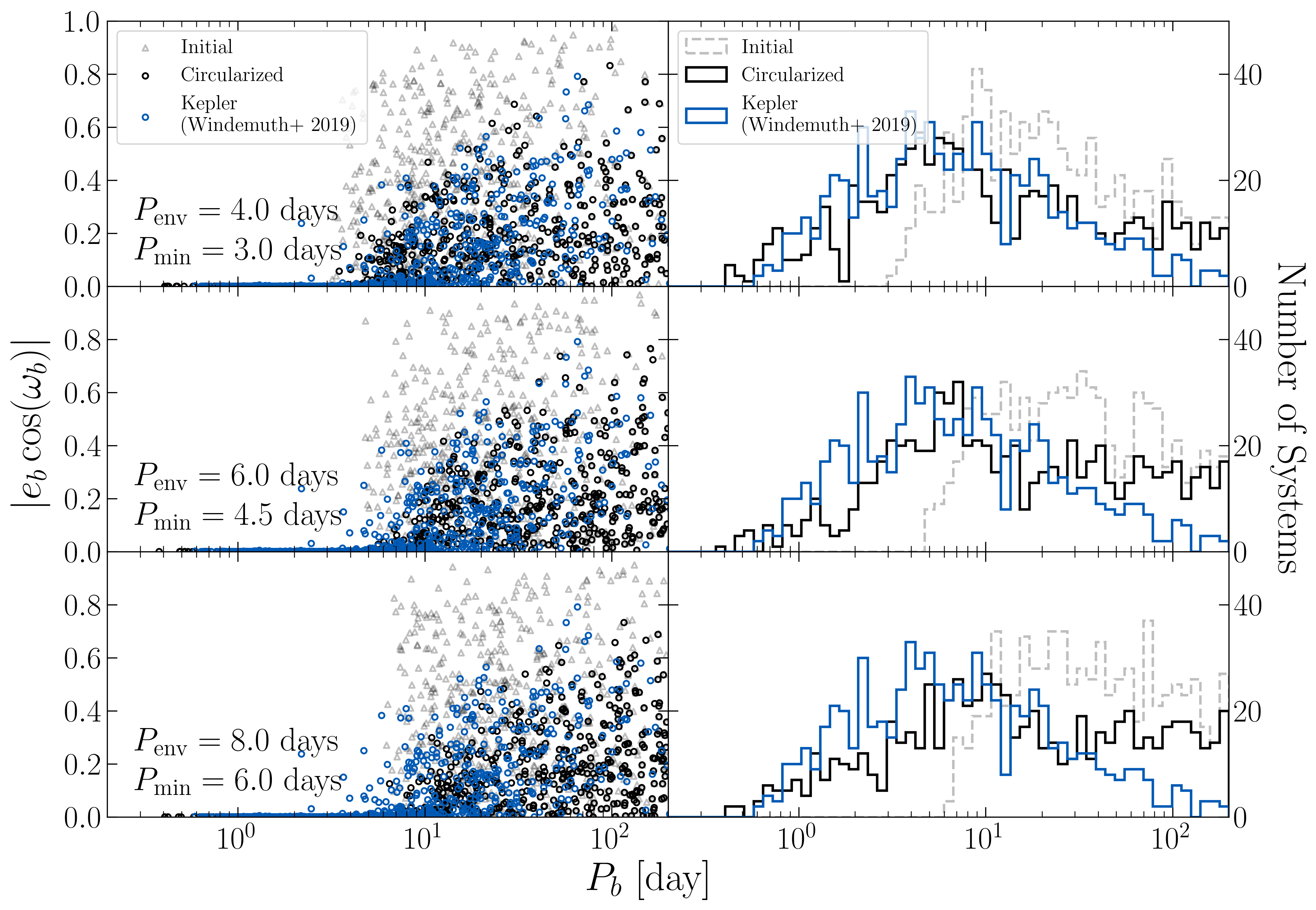}
    \caption{
    Our simulations of eclipsing binaries that have undergone tidal decay (black), compared to the \textit{Kepler} eclipsing binaries from \cite{Windemuth+(2019)} (blue), for the $P_{\rm env}$ (eq.~\ref{eq:t_c}) and $P_{\rm min}$ (eq.~\ref{eq:P_draw}) values indicated.  We also display a mock observation of our initial, un-circularized binaries for reference (gray).
    \textit{Left panels}: Projected eccentricity, perpendicular to the observer's line-of-sight $|e_b \cos (\omega_b)|$, verses orbital period $P_b$.
    \textit{Right panels}: Histograms of the simulated versus observed binary periods.  The canonical parameters $(P_{\rm env}, P_{\rm min}) = (4.0, 3.0)$ day agree best with the observations, but larger values of $(P_{\rm env}, P_{\rm circ})$ are explored, if the low-mass host stars of circumbinary planets undergo stronger tidal decay (see text for discussion).
    \label{fig:bin}
    }
\end{figure*}

Our framework for decaying the orbits of binaries follows \cite{Zanazzi(2022)}, which phenomenologically constrained the efficiency of tidal dissipation using the eccentricity distributions of eclipsing \citep[e.g.][]{Windemuth+(2019), JustesenAlbrecht(2021)} and spectroscopic \citep[e.g.][]{MeibomMathieu(2005), Milliman+(2014)} binaries.  Initial eccentricities $e_{b,0}$ are drawn from a Beta distribution with $a=1.75$ and $b=2.01$ \citep[e.g.][]{Price-Whelan+(2020)}. 
 
We take the primary's mass to be solar ($M_A = 1 \, M_\odot$), with the mass-ratio $q = M_B/M_A \le 1$ drawn from a linear distribution $\propto q$, since short-period binaries have an excess of twins \citep[e.g.][]{LucyRicco(1979), Tokivinin(2000), PinsonneaultStanek(2006), KobulnickyFryer(2007), MoeDiStefano(2017), Kounkel+(2019)}.  The primary's radius is solar ($R_A = R_\odot$), while the secondary's radius $R_B = R_\odot (M_B/M_\odot)^{0.8}$ \citep[e.g.][]{Kippenhahn+(2013)}.  

We circularize our binaries using a constant time-lag model, assuming the rotation rate of the primary is pseudo-synchronous \citep[e.g.][]{Hut(1981),Leconte+(2010)}. As in \cite{Zanazzi(2022)}, we integrate the equations
\begin{align}
    \frac{1}{e} \frac{\der e}{\der t} &= \frac{q(1+q)}{t_c} F_e(e), \\
    \frac{1}{P_{\rm orb}} \frac{\der P_{\rm orb}}{\der t} &= \frac{3 q(1+q)}{2 t_c} F_a(e),
\end{align}
where the functions
\begin{align}
    F_e(e) &= \frac{\Omega_e(e) N(e)}{\Omega(e)} - \frac{18}{11} N_e(e), \\
    F_a(e) &= \frac{4}{11} \left[ \frac{N^2(e)}{\Omega(e)} - N_a(e) \right], \\
    N(e) &= \frac{1 + \frac{15}{2} e^2 + \frac{45}{8} e^4 + \frac{5}{16} e^6}{(1-e^2)^6}, \\
    \Omega(e) &= \frac{1 + 3 e^2 + \frac{3}{8} e^4}{(1-e^2)^{9/2}}, \\
    N_e(e) &= \frac{1 + \frac{15}{4} e^2 + \frac{15}{8} e^4 + \frac{5}{64} e^6}{(1-e^2)^{13/2}}, \\
    \Omega_e(e) &= \frac{1 + \frac{3}{2} e^2 + \frac{1}{8} e^4}{(1 - e^2)^5}, \\
    N_a(e) &= \frac{1 + \frac{31}{2} e^2 + \frac{255}{8} e^4 + \frac{185}{16} e^6 + \frac{25}{64} e^8}{(1 - e^2)^{15/2}},
\end{align}
parameterizing the circularization timescale $t_c$ via 
\begin{equation}
    t_c=0.3\text{ Gyr}\left(\frac{P_{\rm orb}}{P_{\rm env}}\right)^\eta.
    \label{eq:t_c}
\end{equation}

The parameter $P_{\rm env}$ captures the `eccentricity envelope' of the eclipsing binary distribution \citep{Zanazzi(2022)}.  Its value encodes the orbital period below which binaries have predominantly circular orbits, with eccentricities not exceeding a few percent (see Fig.~\ref{fig:bin} left panels, below).  For a combined sample of 524 \textit{Kepler} and TESS eclipsing binaries, \cite{Zanazzi(2022)} found $P_{\rm env} \approx 3$ days, but did examine how the effective temperature of the host may affect $P_{\rm env}$.
\cite{Bashi+(2023)} found for 10,0325 \textit{Gaia} spectroscopic binaries, that $P_{\rm env}$ rose linearly from $\sim$3 to $\sim$7 days as the primary's effective temperature $T_{\rm eff}$ decreased from $\sim$6800 to $\sim$5700 K.  Similarly, \cite{IJspeert+(2024)} found from their sample of 14,573 TESS eclipsing binaries that $P_{\rm env}$ rose as $T_{\rm eff}$ decreased, but only from $\sim$2 to $\sim$3 days as $T_{\rm eff}$ decreased from $\sim$10,000 to $\sim$6000 K.  We assume a canonical envelope period $P_{\rm env} = 4 \ {\rm days}$, but explore longer values of $P_{\rm env} = 6 \ {\rm days}$ and 8 days, since the host stars of circumbinary planets typically have sub-solar masses (Table~\ref{tbl:obs}).  The value of $\eta$ encodes how sharply the tidal dissipation rate depends on orbital separation, and varies for different theories (see \citealt{Zanazzi(2022)} for a discussion).  We vary $\eta$ between values of 2, 4.5, and 7 for our calculations, with a canonical value of $\eta=4.5$.  

For our Monte-Carlo simulations, we draw initial orbital periods using
\begin{equation}
    \ln P_{\rm orb} = (1 - \lambda) \ln P_{\rm min} + \lambda \ln P_{\rm max},
    \label{eq:P_draw}
\end{equation}
with $\lambda$ drawn from a linear distribution over the interval $[0, 1]$, and $P_{\rm max} = 200$ days.  We increase $P_{\rm min}$ with $P_{\rm env}$, taking $P_{\rm min} = 3, 4.5, 6$ days when $P_{\rm env} = 4, 6, 8$ days, respectively, so that most binaries interior to $P_{\rm env}$ will undergo significant tidal decay.  

We integrate the system to an age drawn from a uniform distribution between 1 and 10 Gyr, using the \texttt{LSODA} method of \texttt{solve\_ivp} from \texttt{scipy.integrate} with default tolerances.  After circularization, we remove any binaries with semi-major axis $a_b$ which satisfy
\begin{equation}
    \frac{R_A}{a_b(1-e_b)}\ \geq\ \frac{0.44}{q^{0.13}(1+q)^{0.2}},
\end{equation}
as these systems will undergo Roche lobe overflow \citep{Eggleton(1983)}.  Roche unstable systems comprise only 0.24\% of the total population (combining $\eta = 2$, $4.5$, and $7$ runs).

After circularization, we determine the fraction of binaries which are eclipsing relative to the sky plane of a distant observer.  We draw longitudes of ascending nodes $\Omega_b$ and arguments of pericenter $\omega_b$ from uniform distributions bounded by $0$ and $2\pi$. The binary inclination $i_b$ is drawn from a uniform-in-cosine distribution from $0^\circ$ to $30^\circ$. 

If $i_b$ satisfies
\begin{equation}
    \sin i_b \le \frac{R_A + R_B}{a_b} \left( \frac{1 - |e_b \cos \omega_b|}{1 - e_b^2} \right),
    \label{eq:eclipse}
\end{equation}
we classify the binary as eclipsing \citep{Winn(2010)}.  We truncate at $i_b = 30^\circ$, since for a $i_b > 30^\circ$ equal-mass, circular, solar binary to eclipse, its orbital period must be shorter than 17.5 hours.

Our simulated populations have the best agreement with the \textit{Kepler} eclipsing binaries when $(P_{\rm env}, P_{\rm min}) = (4, 3)$ day (Fig.~\ref{fig:bin}).  For reference, we display a population of ``uncircularized'' eclipsing binaries, by assigning binaries orbital elements without letting their orbits undergo tidal decay, and checking if $i_b$ satisfies condition~\eqref{eq:eclipse}.  Clearly, all binaries interior to $P_{\rm min}$ tidally migrated to their observed orbits.  All models over-predict the number of systems exterior to $P_b \gtrsim 50$ days, which exceeds the binary host period for every transiting circumbinary planet (Fig.~\ref{fig:motiv}).  

We find that $P_{\rm env}$ and $P_{\rm min}$ both set the maximum distance at which orbital decay can hide transiting circumbinary planets. When we increase $P_{\rm env}$ without increasing $P_{\rm min}$, binaries with initial eccentricities $e_{b,0} \lesssim 0.2$ shorten their orbital periods by a paltry amount.  Transiting planets are then disproportionately detected when the host period lies near $P_{\rm min}$, because observational selection effects favor binaries with stagnant, short orbits.  Our $P_{\rm min}$ values for $P_{\rm env}$ are chosen so the orbital period distribution of each tidal model resembles the period distribution of the \textit{Kepler} eclipsing binaries (Fig.~\ref{fig:bin} right panels).  Our tidal shrinkage mechanism comes with the caveat that it depends on the unknown initial eccentricity and orbital period distributions of binaries.

\subsection{Circumbinary Planet Distribution} \label{subsec:planet}

Each binary in our model hosts a circumbinary planet.  We assume planets around binaries have similar occurrence rates as those around single stars, which go as power-laws with orbital period \citep[e.g.][]{Cumming+(2008), Bryan+(2016), Petigura+(2018), Fernandes+(2019), Dattilo+(2023)}.  We draw the planet period $P_p$ from a power-law distribution $\propto P_p^{\alpha}$, varying $\alpha$ between values of $-0.5$, $0$, $0.5$, and $1$.  We choose a canonical value $\alpha=0.5$ 
corresponding to warm sub-Saturns \citep[e.g.][]{Petigura+(2018)}, since most circumbinary planets have radii between $\sim$0.3--0.8 $R_{\rm Jup}$ \citep[see e.g. Table 1 of][]{Li+(2016)}.  Our power-law is bounded below by orbital stability $P_{p,\rm min}=P_{\rm HW}$, and take the maximum period $P_{p, \rm max} = 100 \ P_{\rm HW}$.  Here,
\begin{equation}
    \begin{split}
    \frac{a_{\rm HW}}{a_b} =\ & 1.6+4.12\mu+5.1e_b-4.27\mu e_b \\
                              & -2.22e_b^2-5.09\mu^2+4.61e_b^2\mu^2,
    \end{split}
\end{equation}
and $P_{\rm HW}$ is related to $a_{\rm HW}$ by Kepler's third law, with $\mu = q/(1+q)$ the modified mass ratio \citep{HolmanWiegert(1999)}.  We neglect modifications to $a_{\rm HW}$ due to resonances \citep[e.g.][]{Mardling(2013), LamKipping(2018), Quarles+(2018), MartinFitzmaurice(2022)}, additional islands of stability \citep[e.g.][]{LangfordWeiss(2023)}, and large mutual inclinations \citep[e.g.][]{Chen+(2020)}.

Because $P_{\rm HW}$ decreases as the binary orbit shrinks, when the planets form relative to when the orbit decays affects the planets' period distribution.  We consider multiple prescriptions to test the effect of shrinkage on the detection of transiting planets.  For our first, control case, we let planets form around the \textit{Kepler} eclipsing binaries observed in \cite{Windemuth+(2019)}, assigning binary eccentricities $e_b$ from the measured $e_b \cos \omega_b$ value by drawing $\omega_b$ uniformly between $0$ and $\cos^{-1}(e_b \cos \omega_b)$, and setting $e_b = (e_b \cos \omega_b)/\cos \omega_b$.  The $P_{\rm HW}$ is set by the binary's current orbit.  In our second case, we circularize binaries as discussed in Section~\ref{subsec:binary}, and assume planets form after the binary shrinks, populating planets exterior to the stability boundary $P_{\rm HW}$ after tidal decay.  For our third case,  planets form before, not after, the binary orbit decays.  The stability boundary $P_{\rm HW}$ is set by the binary's initial orbit, leaving planets around short-period binaries stranded at larger separations after binary orbital periods shorten.  Each calculation includes $10^5$ circumbinary planets.

We assume the planet's orbit forms slightly misaligned with the binary's orbital plane.  The initial mutual inclination between the orbit of the planet and binary $i_m$ is drawn from a Rayleigh distribution, varying scale parameter values between $\sigma_m = 0.3^\circ$, $1^\circ$, and $3^\circ$, with $\sigma_m=1^\circ$ our canonical value.  In the frame where the binary's pericenter direction lies on the $x$-axis, and orbit normal defines the $z$-axis, we draw the planet's initial longitude of ascending node $\Omega_m$ and true anomaly $f_m$ from uniform distributions between $0$ and $2\pi$.  We then calculate the planet's orbital elements $\Omega_p$ and $i_p$ in the sky plane. Details of this transformation are contained within Appendix~\ref{apdx:sky}.

For simplicity, we assume the planet has no radius ($R_p=0$) or mass ($M_p=0$), and that its orbit is initially circular ($e_p=0$).  Sampling initial $e_p$ values from a Rayleigh distribution with a scale parameter $\sigma_p=0.05$, and setting the mass and radius of the planet equal to that of Neptune, did little to change our results.

\subsection{Planetary Transit Detection}\label{subsec:transit}

\begin{figure}
    \includegraphics[width=\linewidth]{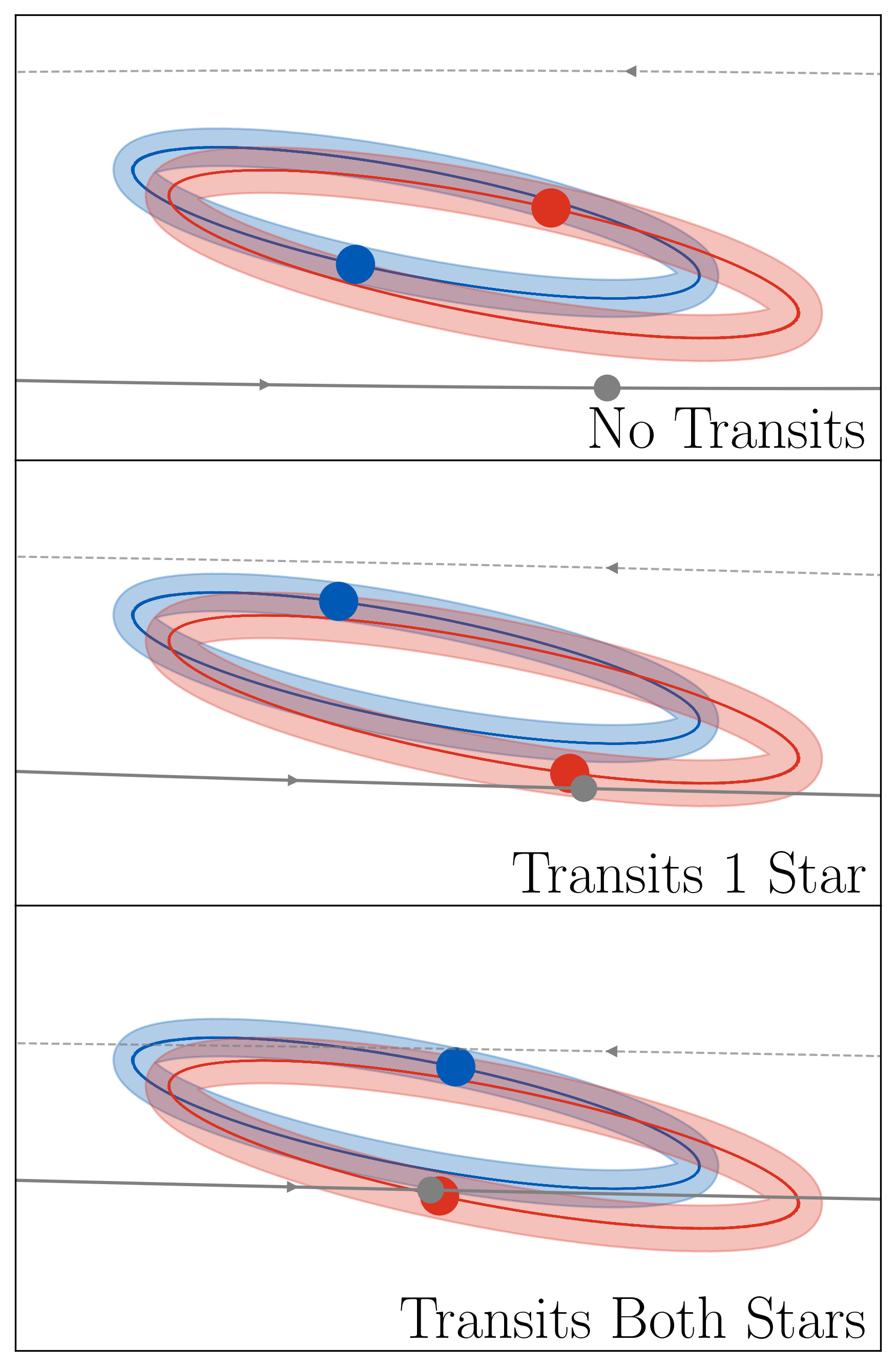}
    \caption{
    Visualization of our transit checking algorithm.  The orbits of two binaries (blue and red) are projected onto the sky plane of a distant observer.  The planet's orbit (gray) is classified as transiting if the portion that lies between the binary and observer (solid line) intersects both stellar orbits (bottom panel).  For illustrative clarity, our example displays the orbits of non-eclipsing binaries.
    \label{fig:donut}}
\end{figure}

For every system in the Monte Carlo simulation, we integrate the orbit of the binary and planet using the N-body integrator \texttt{Rebound}, and use a geometrical algorithm to test if each system allows any planetary transits.  We use the \texttt{IAS15} integrator over a time period of 8 years, slightly longer than the combined \textit{Kepler} and \textit{K2} mission baselines of $\sim$4 and $\sim$3 years, respectively.  If a planet transits the binary's orbital plane, the system's orbital properties are recorded. 

Rather than exhaustively checking if the planet itself crosses the paths of one of the stars in the binary and the observer's line-of-sight, at each time-step of our N-body integration, we smear the trajectory of the planet and projected area of each binary component, and determine if they intersect (Fig.~\ref{fig:donut}).  The planet's trajectory is the half of the elliptical orbit which lies in the positive $z$ direction of the sky plane, assuming transits occur when the planet lies between the binary's orbit and the observer's line-of-sight, while each star in the binary is taken to be a pipe surface with radius $R_A$ or $R_B$.  If the planet's trajectory crosses any of the pipe surfaces, the star which the planet transited is recorded. Only 1.36\% of transiting planets transit one, and not both, stellar components of the binary.

\begin{figure}
    \includegraphics[width=\linewidth]{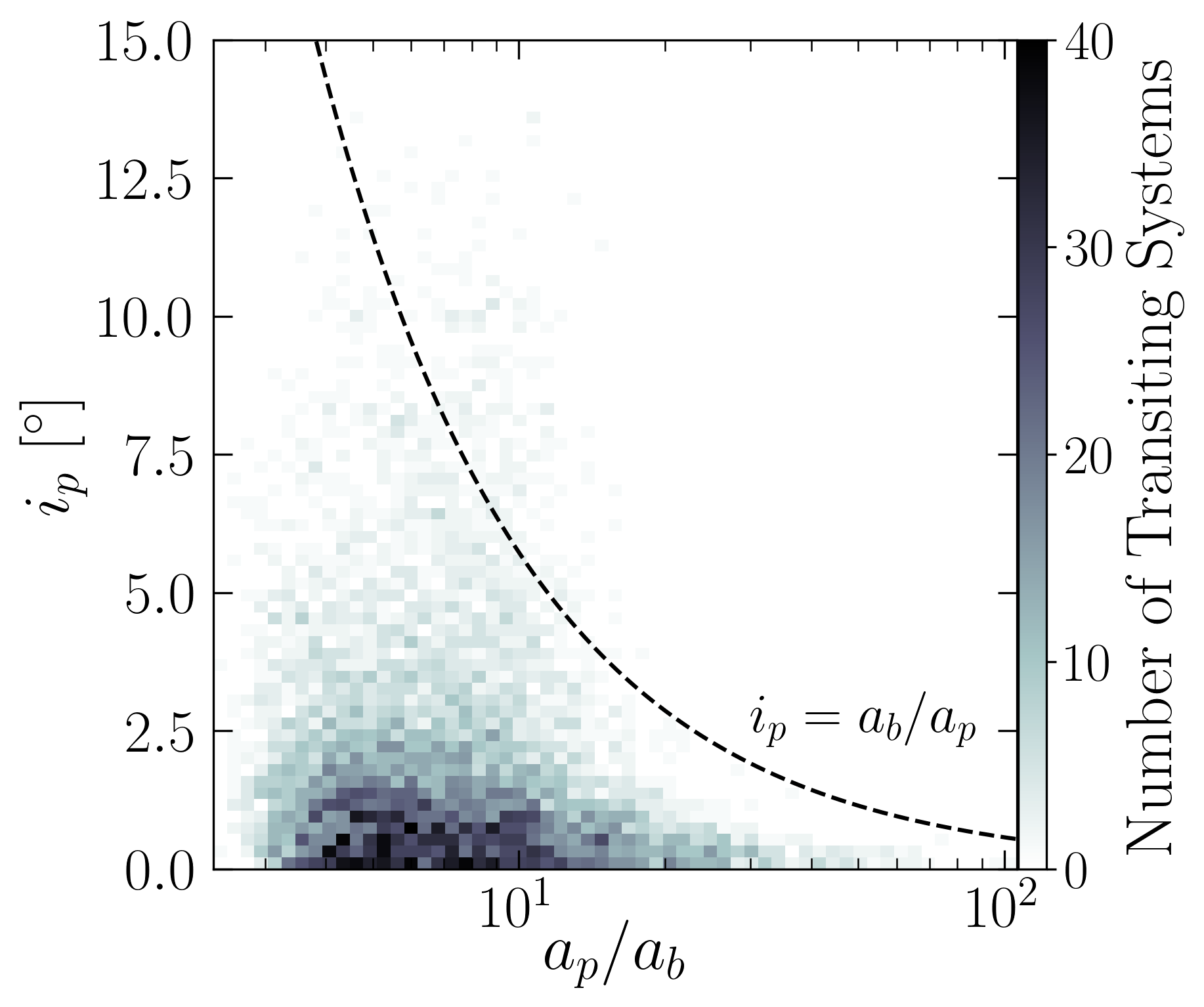}
    \caption{Transiting circumbinary planets for our canonical population synthesis model, versus the initial planet inclination projected onto the sky plane $i_p$, and the planet separation $a_p/a_b$.  The dashed line shows the limiting $i_p$ above which we do not expect many transiting planets (eq.~\ref{eq:ip_limit}); systems detected above $i_p = a_b/a_p$ nodally-precess into the observer's line-of-sight.  
    \label{fig:transit_check}}
\end{figure}

As the planet's separation increases, the range of planetary inclinations $i_p$ where the width of the planet's projected orbit $Z_p \approx 2 a_p \sin i_p$ is smaller than that of the binary $Z_b \approx 2 a_b \sin i_b$ decreases.  Transits occur only if $Z_p \lesssim Z_b$, which requires inclinations below
\begin{equation}
    i_p \lesssim \frac{a_b}{a_p} \sin i_b \lesssim \frac{a_b}{a_p},
    \label{eq:ip_limit}
\end{equation}
assuming $i_p \ll 1$.  Figure~\ref{fig:transit_check} plots the number of detected systems from our canonical tidal decay population synthesis (see Sec.~\ref{sec:results} below for details).  As expected, most systems are detected only if the planet's initial $i_p$ lies well below $a_b/a_p$.  Systems detected with $i_p$ values above $a_b/a_p$ change $Z_p$ over the course of the simulation, due to nodal precession \citep[e.g.][]{MartinTriaud(2015)}.  Our simulations reproduce \cite{MartinTriaud(2014)}, who found transiting planets often orbit non-eclipsing binaries: for 

our canonical population synthesis which formed planets before tidal decay, we detected 599 planets orbiting eclipsing binaries, and 999 planets orbiting non-eclipsing binaries.

\section{Results} \label{sec:results}

Here, we compare the distributions of transiting planets orbiting binaries from our Monte Carlo calculations (Sec.~\ref{sec:model}) to those observed (Table~\ref{tbl:obs}).  We run 10 separate calculations, either drawing binary properties from those observed in \textit{Kepler} eclipsing binaries, or from a binary orbital decay population synthesis.  We vary the parameters $\eta$, $\alpha$, and $\sigma_m$; see Section~\ref{sec:model} for definitions.

\subsection{Transiting Circumbinary Planet Distributions}

\begin{figure}
    \includegraphics[width=\linewidth]{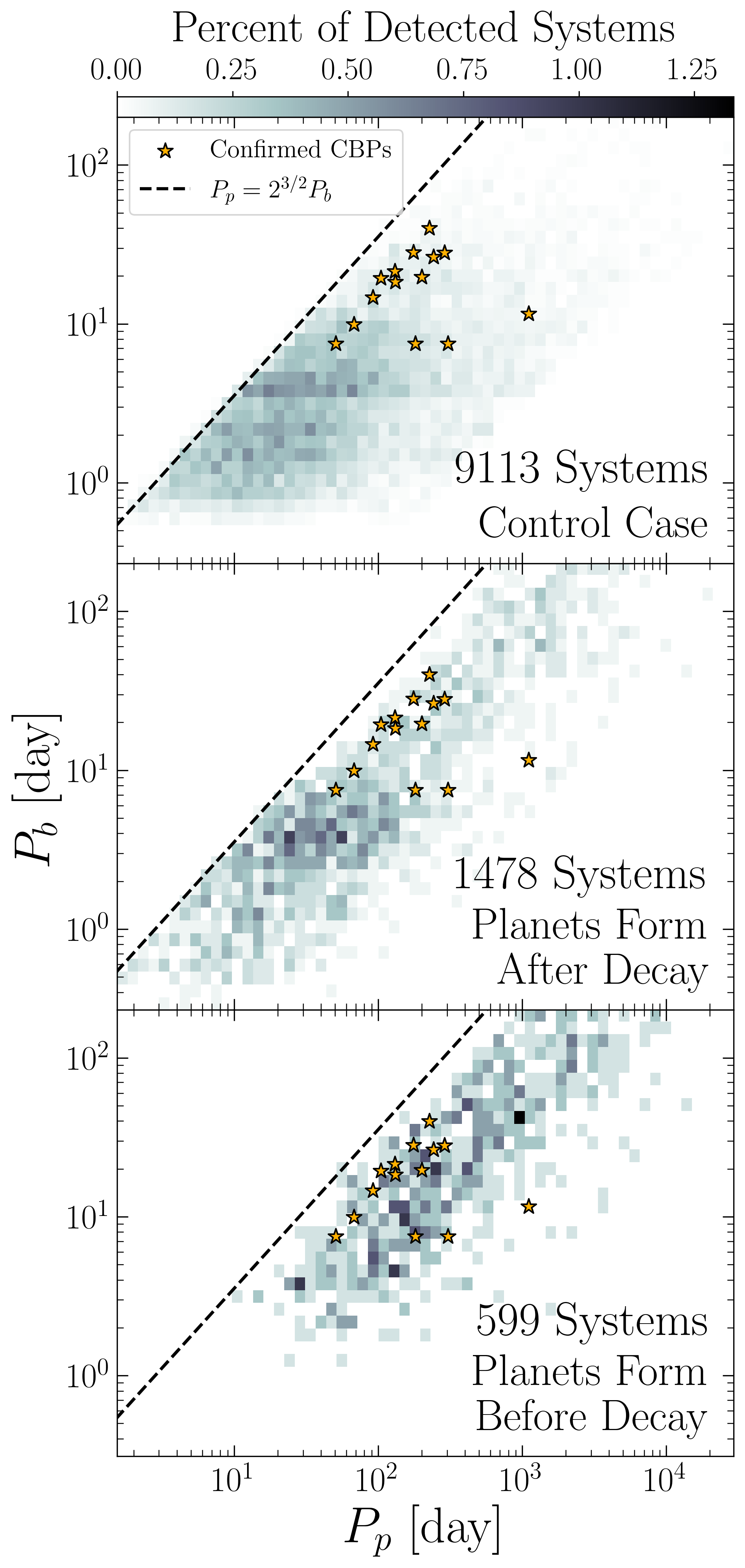}
    \caption{
    Synthetic populations of transiting circumbinary planets.  Colors indicate the percent of transiting planets detected at a given binary $P_b$ and planetary $P_p$ orbital period, with stars displaying observed systems for reference (Table~\ref{tbl:obs}).  Our control case, placing planets around \textit{Kepler} eclipsing binaries (top panel) is compared to the binary population synthesis, placing planets up to the stability limit $P_{\rm HW}$ after (middle panel), versus before (bottom panel), the binary orbit decays.  The dashed line displays the approximate stability limit, with the total number of transiting systems indicated on each plot.  Binary decay following planet formation conceals transiting planets orbiting short $P_b$ hosts.
    \label{fig:2dhist_control}}
\end{figure}

\begin{figure}
    \includegraphics[width=\linewidth]{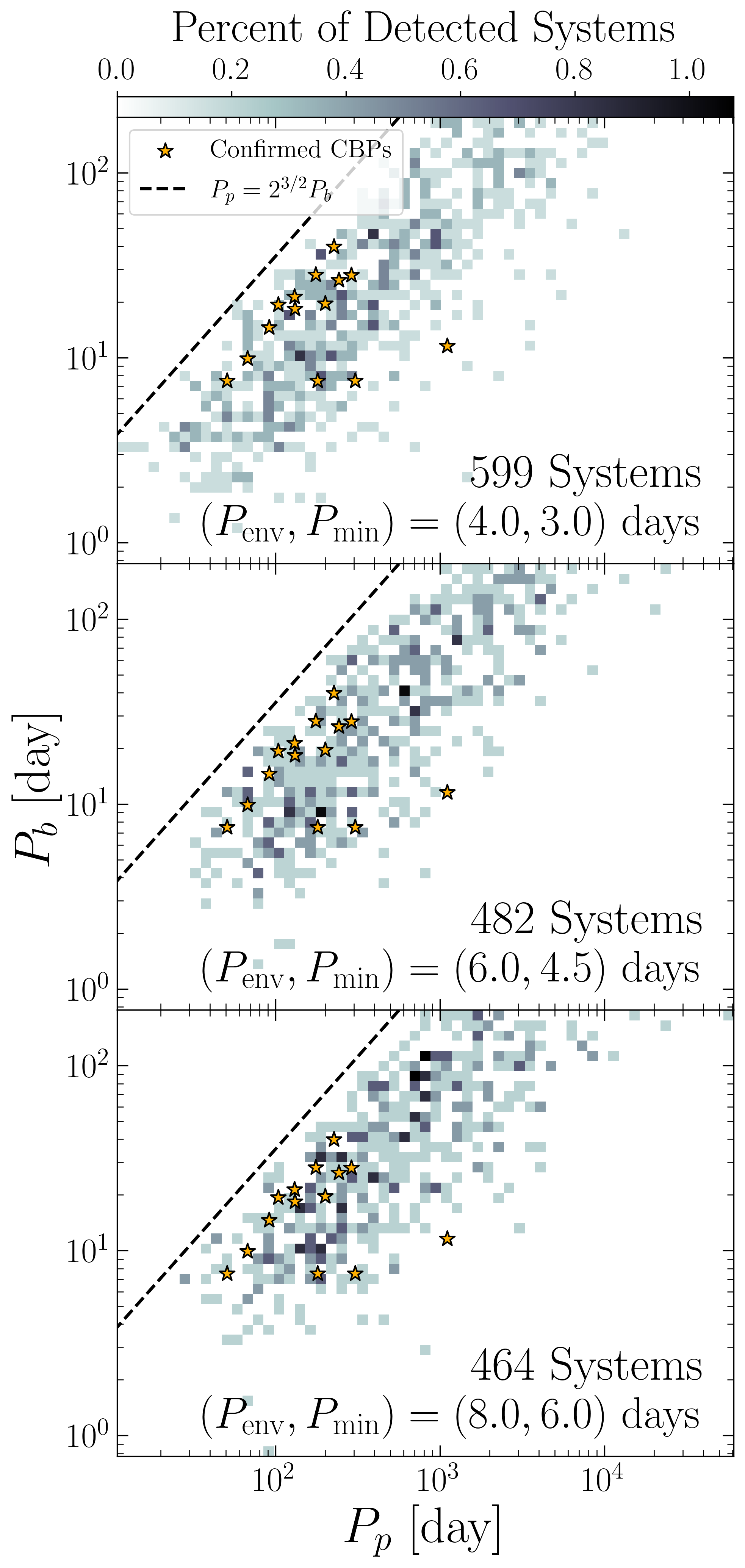}
    \caption{
    Same as the bottom panel of Figure~\ref{fig:2dhist_control}, varying $P_{\rm env}$ (eq.~\ref{eq:t_c}) and $P_{\rm min}$ (eq.~\ref{eq:P_draw}) for our tidal decay model.  Increasing $(P_{\rm env}, P_{\rm min})$ also increases the binary period of transiting planet hosts.
    \label{fig:2dhist_penv}}
\end{figure}

\begin{figure*}
    \includegraphics[width=\linewidth]{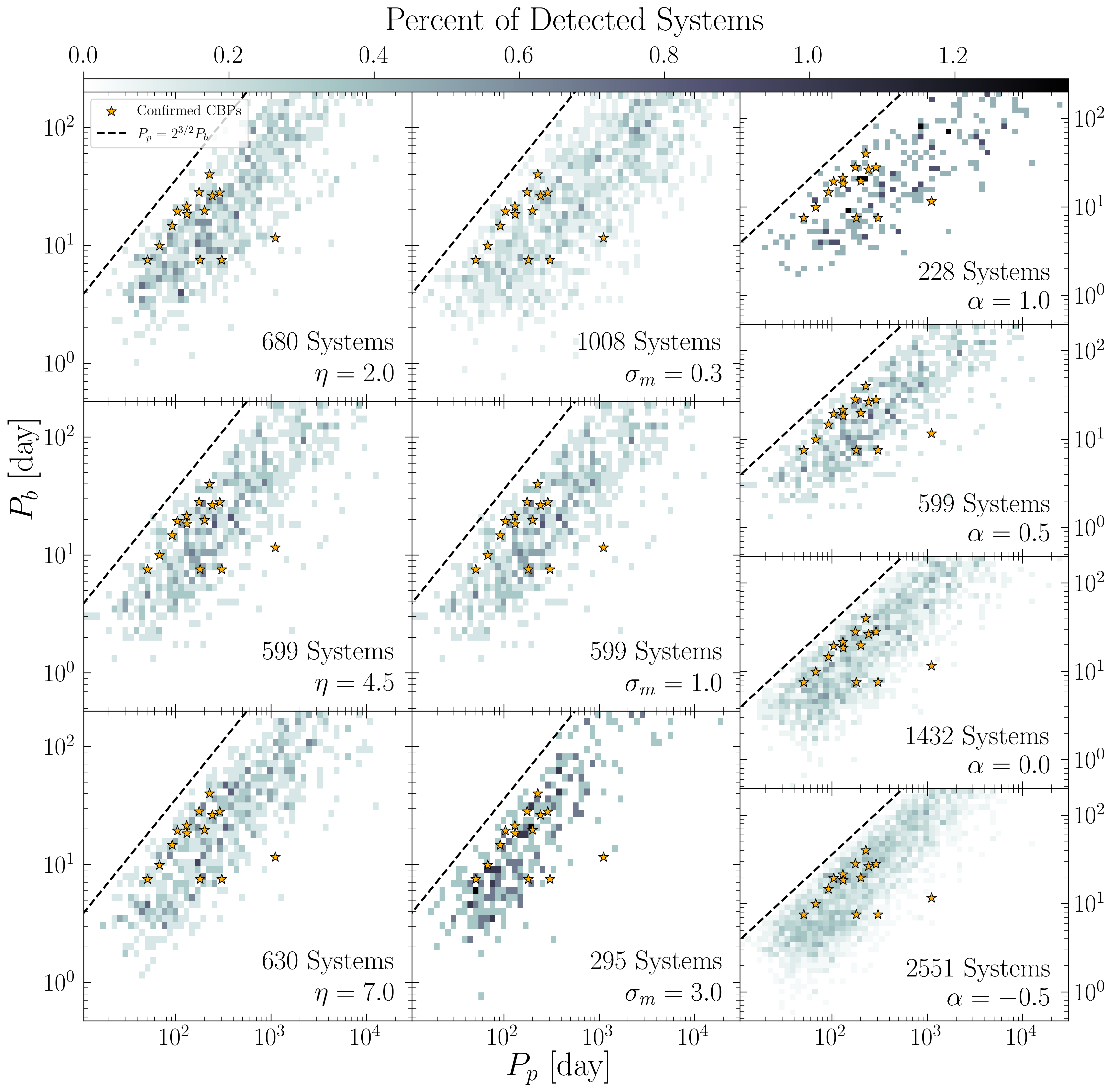}
    \caption{Same as the bottom panel of Figure~\ref{fig:2dhist_control}, varying $\eta$, $\sigma_m$, and $\alpha$. Our results are relatively insensitive to variations in $\eta$, $\sigma_m$, and $\alpha$. \label{fig:2dhist_params}}
\end{figure*}

Our calculations suggest that the orbital decay of binaries can hide circumbinary planets, when the binary orbital period is sufficiently short.  Figure~\ref{fig:2dhist_control} compares our control case, placing planets around \textit{Kepler} eclipsing binaries, to Monte Carlo simulations placing planets exterior to the stability boundary after, versus before, the binary period shortens.  If planets are placed around the binary after its orbit decays, we expect most planets to be detected around sub-seven-day binary hosts, in 
disagreement 
with the observed circumbinary planets. However, if the planet forms before the binary orbit decays, a precipitous drop in planets detected around short-period binary hosts occurs.  Nonetheless, some `bleeding' of the circumbinary planet host distribution is seen, where planets remain detectable around sub-seven-day period host binaries, down to about $\sim$3 days.

If we increase the amount of tidal shrinkage, this bleeding can be cauterized.  When $P_{\rm env}$ and $P_{\rm min}$ are raised to higher values, tidal circularization and decay occurs out to longer binary periods, causing more circumbinary planets to be concealed (Fig.~\ref{fig:2dhist_penv}).  Our models with $(P_{\rm env}, P_{\rm min}) = (6.0, 4.5)$ days and $(8.0, 6.0)$ days seem to better agree with the observed circumbinary systems, with transiting planets not detected at periods shorter than $\sim$4 and $\sim$7 days, respectively.  The statistical significance of each model's bleeding is discussed more in Section~\ref{subsec:minp} below.

The host period $P_b$ distributions of transiting circumbinary planets do not seem to depend strongly on $\eta$, $\sigma_m$, or $\alpha$ (Fig.~\ref{fig:2dhist_params}).  Increasing $\eta$ weakens the rate of tidal dissipation at large binary separations \citep{Zanazzi(2022)}, causing fewer binaries to circularize, with more planets detected near the stability boundary.  Increasing $\sigma_m$ decreases the planets detected at long planet periods $P_p$, since planets with large mutual inclinations have low detection probabilities.  Decreasing $\alpha$ increases the number of planets that form, and hence are detected, at short $P_p$.

\subsection{Shortest Host Binary Period} \label{subsec:minp}

\begin{figure}
    \includegraphics[width=\linewidth]{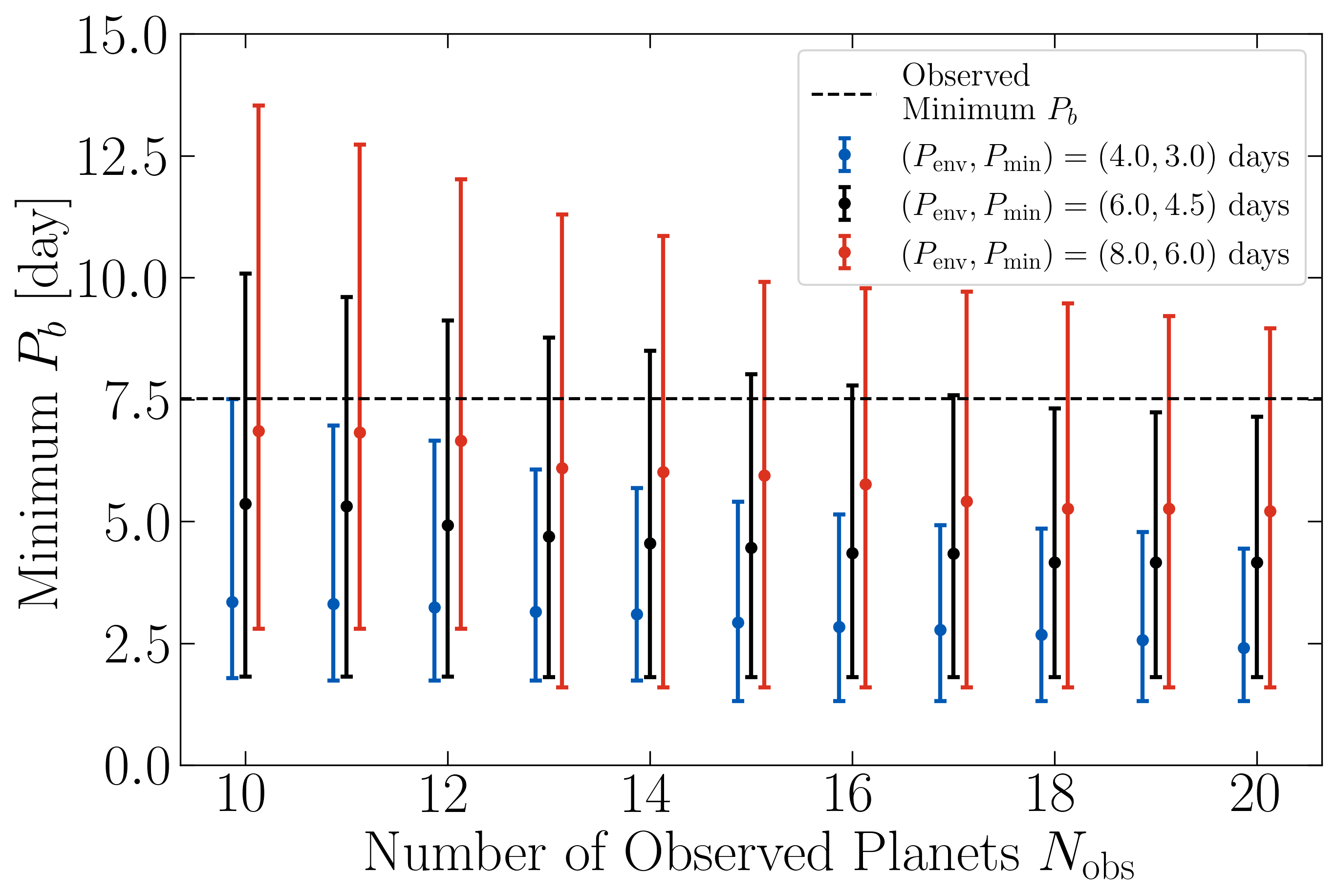}
    \caption{The minimum period of binary hosts $P_b$ after observing $N_{\rm obs}$ circumbinary planets, drawing systems from the distributions displayed in Figure~\ref{fig:2dhist_penv}.  Increasing $(P_{\rm env}, P_{\rm min})$ also increases the minimum $P_b$, and as $N_{\rm obs}$ increases, the minimum $P_b$ decreases.
    \label{fig:minp_penv}}
\end{figure}

Here, we draw a subsample of $N_{\rm obs} \in \{10, 11, 12, \dots, 20\}$ `observed' planets transiting circumbinary planet hosts from our population syntheses, and determine the range of minimum host periods $P_b$ from this sample after drawing $N_{\rm obs}$ a total of $10^5$ times.  

Figure~\ref{fig:minp_penv} displays the 90\% confidence intervals for our control case, compared to our Monte Carlo models which shrink binaries to larger separations by increasing $P_{\rm env}$ and $P_{\rm min}$.  For our canonical binary-decay-after-planet-formation model ($P_{\rm env}, P_{\rm min} = 4.0, 3.0$ days), the minimum $P_b$ has a median value of 3.2 days, and an upper limit of 5.8 days, when $N_{\rm obs} = 13$.  These values for the minimum $P_b$ do not quite agree with the 7.4 day minimum period observed (Table~\ref{tbl:obs}).  However, pushing $(P_{\rm env}, P_{\rm min})$ to larger values raises the minimum $P_b$, and for our models with $(P_{\rm env}, P_{\rm min}) = (6.0, 4.5)$ days and $(8.0, 6.0)$ days, our confidence intervals encompass \textit{Kepler} 47's observed 7.4 day period (Fig.~\ref{fig:minp_penv}).  We find that the tidal decay of binaries could conceal transiting planets orbiting sub-seven-day-period hosts, although the requisite magnitude of tidal dissipation might exceed phenomenological constraints from the eccentricity distributions of short-period binaries \citep[e.g.][]{Zanazzi(2022), Bashi+(2023), IJspeert+(2024)}.

\section{Summary and Discussion} \label{sec:disc}

As a stellar binary circularizes, its orbital separation shrinks, decreasing the area of the orbit projected onto the observer's sky plane.  We have investigated to what extent and under what circumstances this orbital shrinkage can prevent circumbinary planets from transiting their host stars, to see whether we can reproduce the lack of transit-hosting binaries with periods $P_b < 7$ days (Fig.~\ref{fig:bias}).  Our Monte Carlo calculation uses a phenomenological model for tidal dissipation, with free parameters adjusted to reproduce, to varying extents, the observed $P_b$ and eccentricity $e_b$ distributions of eclipsing binaries (Fig.~\ref{fig:bin}).  Planets are assumed to form with small inclinations relative to their host binaries, and labeled transiting if over an 8-year N-body integration they cross the binary orbit  (Fig.~\ref{fig:donut}).  If circumbinary planets form after binaries shrink, tides introduce no observational selection effect on the transiting statistics (Fig.~\ref{fig:2dhist_control} top and middle panels).  By contrast, planet formation before orbital decay may, depending on the binary orbital parameters at the time of formation, and the efficiency of dissipation, result in far fewer transiting planets around $P_b < 7 $ day binaries (Fig.~\ref{fig:2dhist_control} bottom panel). 
Our population synthesis can reproduce the deficit of transiting planets at $P_b < 7$ days, provided binaries tidally circularize and decay out to $P_b \sim$ 6--8 days, which they might for lower-mass stars (Figs.~\ref{fig:2dhist_penv} and \ref{fig:minp_penv}; see \citealt{Bashi+(2023)} for evidence supporting 6--8 days, but \citealt{IJspeert+(2024)} for evidence against).  Our results are sensitive to the binary eccentricity and orbital period distributions at formation; reproducing the observed transit statistics requires that all current-epoch $< 6$ day binaries were born with periods $> 6$ days and substantial eccentricities.

To reiterate the last point: the scenario explored in this paper requires that no binaries form with nearly-circular orbits and $<$ 6 day periods, or at least not with close-in circumbinary planets. This conclusion is unsatisfying because it almost just restates the original problem of why circumbinary planets are not observed around $<$ 7 day binaries. It also seems an unreasonable demand. For example, a nearly-circular, tight stellar binary could form by disk torques (e.g.~\citealt{MoeKratter(2018)}). The same circumbinary disk that drove binary migration could also spawn a close-in transiting planet.

Despite these conceptual problems, the observational data do not obviously rule out the possibility that $P_b < 6$ day binaries arrived at their orbits predominantly from post-formation tidal shrinkage.
Pre-main-sequence binaries with $P_b < 6$ days display a `waterfall' shape in period-eccentricity space attributed to tidal circularization \citep[e.g.][]{Melo+(2001), MeibomMathieu(2005)}, implying they may have formed with higher eccentricities.  Numerous tidal theories have been proposed to circularize binaries at such young ages \citep[e.g.][]{Zahn(1989), ZanazziWu(2021), Terquem(2021), Barker(2022)}.  The `circularization period' marking where the period-eccentricity waterfall transitions from eccentric to circular orbits also grows with age \citep[e.g.][]{MeibomMathieu(2005), Milliman+(2014), Zanazzi(2022)}, lending further support to this feature being caused by tidal circularization.  We can also examine binaries hosting planet-forming disks.  Of the dozens of protoplanetary and debris circumbinary disks identified in \cite{Czekala+(2019)}, nearly all have periods $\gtrsim 10$ days and eccentricities $\gtrsim$ 0.1. Two exceptions include the circular, 2.4 day and 3.9 day period binaries V4046 Sgr and CoRoT 2239, respectively. Perhaps these binaries arrived at their orbits from tidal shrinkage, but if they did, it remains an unsolved problem why they would not host circumbinary planets up to their stability boundaries.

The hypothesis of tidal shrinkage is at least predictive: there should exist a population of nearly co-planar, wide-separation circumbinary planets that do not transit.  Such systems could be detected orbiting eclipsing binaries via eclipse timing variations.  A possible exemplar  is KIC 3853259 (AB)b, a tight 0.28 day period binary that displays eclipse timing variations from a non-transiting, 326 day period, $\sim$10$M_{\rm Jup}$ mass planetary companion \citep{Borkovits+(2016)}.  The detection of more such systems would support tidal shrinkage hiding transiting circumbinary planets.  The scenario predicts more transiting and non-transiting planets to be detected orbiting sub-seven-day period binaries, far from the stability boundary on nearly co-planar orbits.  This architecture differs from planets hidden by the von Zipel-Lidov-Kozai mechanism, which predicts misaligned circumbinary planet orbits \citep[e.g.][]{MunozLai(2015)}. 

\begin{acknowledgments}
We thank Eugene Chiang for his extensive feedback regarding the content of this manuscript, and Mohammad Farhat for pointing out the existence of KIC 3853259 (AB)b.  We also thank Dan Fabrycky, Gongjie Li, and William Welsh for useful discussions.  SM was funded in part by the Berkeley Physics Undergraduate Research Scholars Program (BPURS), and SM and JJZ both received funding through the Heising-Simons Foundation 51 Pegasi b Postdoctoral Fellowship.  This research used the Savio computational cluster resource provided by the Berkeley Research Computing program at the University of California, Berkeley (supported by the UC Berkeley Chancellor, Vice Chancellor for Research, and Chief Information Officer).
\end{acknowledgments}

\appendix

\section{Sky Plane Projection} \label{apdx:sky}

To determine if the planet transits its host binary, we place the planet on an orbit inclined to the binary, then project both the planet and binary orbits onto the sky plane of a distant observer.  We start first with the binary's orbital elements.  We let $\hu$ point in the direction of periapse, $\hl$ define the orbit normal of the binary, and $\hv = \hl \btimes \hu$, so that $(\hu, \hv, \hl)$ form an orthonormal triad.  We let $R_x(\theta)$ denote a rotation about the $x$-axis,
\begin{equation}
    R_x(\theta) = 
    \left(
    \begin{array}{ccc}
    1 & 0 & 0 \\
    0 & \cos \theta & -\sin \theta \\
    0 & \sin \theta & \cos \theta
    \end{array}
    \right),
\end{equation}
and $R_z(\theta)$ denote a rotation about the $z$-axis,
\begin{equation}
    R_z(\theta) = 
    \left(
    \begin{array}{ccc}
    \cos \theta & -\sin \theta & 0 \\
    \sin \theta & \cos \theta & 0 \\
    0 & 0 & 1
    \end{array}
    \right).
\end{equation}
To calculate the components of $(\hu, \hv, \hl)$ in the sky basis $(\hx, \hy, \hz)$ in terms of the binary's argument of pericenter $\omega_b$, inclination $i_b$, and longitude of ascending node $\Omega_b$, we perform a sequence of rotations \citep[e.g.][]{MurrayDermott(1999)}
\begin{equation}
    \left( \begin{array}{c}
    \hu \\
    \hv \\
    \hl
    \end{array} \right)
    = R_z(\Omega_b) \cdot R_x(i_b) \cdot R_z(\omega_b) \cdot 
    \left( \begin{array}{c}
    \hx \\
    \hy \\
    \hz
    \end{array} \right).
\end{equation}
Explicitly in terms of $\hx$, $\hy$, and $\hz$,
\begin{align}
    \hu &= ( \cos \omega_b \cos \Omega_b - \sin \omega_b \cos i_b \sin \Omega_b) \hx
    \nonumber \\
    &+ (\cos \omega_b \sin \Omega_b + \sin \omega_b \cos i_b \cos \Omega_b) \hy
    \nonumber \\
    &+ (\sin \omega_b \sin i_b) \hz, \\
    \hv &= -(\sin \omega_b \cos \Omega_b + \cos \omega_b \cos i_b \sin \Omega_b) \hx 
    \nonumber \\
    &+ ( \cos \omega_b \cos i_b \cos \Omega_b - \sin \omega_b \sin \Omega_b ) \hy
    \nonumber \\
    &+ (\cos \omega_b \sin i_b) \hz, \\
    \hl &= \sin i_b \sin \Omega_b \hx - \sin i_b \cos \Omega_b \hy
    +\cos i_b \hz.
\end{align}

In the binary reference frame ($\hu, \hv, \hl$), we place the planet on an orbit with the argument of pericenter $\omega_m$, inclination $i_m$, and longitude of ascending node $\Omega_m$.  The planet's reference direction $\hx_p$, orbit normal $\hz_p$, and $\hy_p = \hz_p \btimes \hx_p$ are then given by
\begin{equation}
    \left( \begin{array}{c}  \hx_p \\ \hy_p \\ \hz_p \end{array} \right)
    = R_z(-\omega_m) \cdot R_x(-i_m) \cdot R_z(-\Omega_m) \cdot 
    \left( \begin{array}{c} \hu \\ \hv \\ \hl \end{array} \right).
    \label{eq:state_vecs_p}
\end{equation}
Written explicitly in terms of $(\hu, \hv, \hl)$,
\begin{align}
    \hx_p &= (\cos \omega_m \cos \Omega_m - \cos i_m \sin \omega_m \sin \Omega_m) \hu 
    \nonumber \\
    &+ (-\cos \Omega_m \sin \omega_m - \cos i_m \cos \omega_m \sin \Omega_m) \hv 
    \nonumber \\
    &+ (\sin i_m \sin \Omega_m) \hl, 
     \\
    \hy_p &= (\cos i_m \cos \Omega_m \sin \omega_m + \cos \omega_m \sin \Omega_m) \hu 
    \nonumber \\
    &+ (\cos i_m \cos \omega_m \cos \Omega_m - \sin \omega_m \sin \Omega_m) \hv 
    \nonumber \\
    &+ (-\cos \Omega_m \sin i_m) \hl, 
     \\
    \hz_p &= \sin i_m \sin \omega_m \hu + \cos \omega_m \sin i_m \hv + \cos i_m \hl. 
\end{align}

Now we can calculate the planet properties in the sky reference frame ($\hx, \hy, \hz$).  The inclination is given by
\begin{equation}
    i_p = \cos^{-1}\left(\hz_p \bcdot \hz \right).
\end{equation}
We also define a vector which points in the ascending node's direction:
\begin{equation}
    \vn = \hz_p \btimes \hz, \hspace{5mm} \hn = \vn/|\vn|,
\end{equation}
which allows us to calculate the planet's longitude of ascending node
\begin{equation}
    \Omega_p = 
    \left\{ \begin{array}{ll}
    \cos^{-1} \left( \hx_p \bcdot \hn \right) & \text{for } \hy_p \bcdot \hn \ge 0 \\
    2\pi - \cos^{-1}\left( \hx_p \bcdot \hn \right) & \text{for } \hy_p \bcdot \hn < 0
    \end{array} \right. ,
\end{equation}
and argument of pericenter
\begin{equation}
    \omega_p = 
    \left\{ \begin{array}{ll}
    \cos^{-1}(\hn \bcdot \hu) & \hu \bcdot \hz_p \ge 0 \\
    2\pi - \cos^{-1}(\hn \bcdot \hu) & \hu \bcdot \hz_p < 0
    \end{array} \right. .
\end{equation}

\software{All code and data used in this work are publicly available on \href{https://github.com/saahitmog/circumbinary}{GitHub}. We made use of the following publicly available Python modules: \texttt{rebound} \citep{ReinLiu(2012)}, \texttt{astropy} \citep{Astropy}, \texttt{shapely} \citep{Gillies(2024)}, \texttt{matplotlib} \citep{Hunter(2007)}, \texttt{numpy} \citep{Harris(2020)}, \texttt{scipy} \citep{Virtanen(2020)}, and \texttt{pandas} \citep{pandas(2020)}.}

\nocite{*}
\bibliography{ref.bib}{}
\bibliographystyle{aasjournal}

\end{document}